\documentclass[a4paper,12pt]{article}
\usepackage{latexsym}
\usepackage{epsfig}

\def\bg#1{\mbox{\boldmath$#1$}}

\newcommand{\del}{\partial}
\newcommand{\beq}{\begin{eqnarray}}
\newcommand{\eeq}{\end{eqnarray}}
\newcommand{\be}{\begin{eqnarray*}}
\newcommand{\ee}{\end{eqnarray*}}
\newcommand{\bk}{{\bf k}}

\newcommand{\br}{{\bf r}}

\newcommand{\ra}{\rightarrow}

\begin{document}

\centerline{\Large\bf {Scalar Gravitation and Extra Dimensions}}
\vskip 5mm
\centerline{Finn Ravndal\footnote{Invited talk at {\it The Gunnar Nordstr\"om Symposium on
       Theoretical Physics}, Helsinki, August 27 - 30, 2003.}}
\vskip 5mm
\centerline{\it Department of Physics, University of Oslo, N-0316 Oslo, Norway.}

\begin{abstract}

Gunnar Nordstr\"om constructed the first relativistic theory of gravitation formulated in terms of
interactions with a scalar field.
It was an important precursor to Einstein's general theory of relativity a couple of years later. He was
also the first to introduce an extra dimension to our spacetime so that gravitation would be just an aspect 
of electromagnetic interactions in five dimensions. His scalar theory and generalizations thereof are here presented 
in a bit more modern setting.
Extra dimensions are of great interest in physics today and can give scalar gravitational interactions
with similar properties as in Nordstr\"om's original theory.

\end{abstract}

\section{Introduction}

Modern theories of gravitation based on supergravity or superstrings contain in general one or more scalar fields which will
modify at some level the original tensor theory of Einstein. What r\^ole they play in the real world is not at all clear. But
recent developments in cosmology such as the need for a very fast inflationary phase\cite{Kinney} in the early Universe and dark 
energy to explain the slower acceleration much later\cite{Sahni}, are now being discussed in terms of such scalar fields with 
gravitational interactions.

Most of the underlying, fundamental theories require spacetime to have more than the four we know about today. Such extra dimensions must
be microscopic or curled up in some highly non-trivial way. These ideas are so interesting that they have inspired much experimental work
during the last few years\cite{Adelberger}.

It was the Finnish physicist Gunnar Nordstr\"om who 90 years ago was the person who laid the first seeds in this research program
which today has grown to a large, international effort. He was the first to construct a consistent, relativistic theory
of gravitation described by the interactions of a massless, scalar field. Even if this theoretical proposal did not turn out to
be physically correct and was after a couple of years replaced by Einstein's tensor theory\cite{Einstein_1}, it represented an important
milestone in the investigation of gravitational phenomena. Probably more important was his daring idea  to explain
unknown physics in our known spacetime by known physics in a spacetime with extra, compactified dimensions. This was several years
before the more detailed theories of Kaluza\cite{Kaluza} and Klein\cite{Klein} and lies today at the heart of the most fundamental 
theories of Nature ever formulated.

We will here give a short summary of the main ideas behind Nordstr\"om's different contributions in a bit more modern formulation. As
an example of how scalar fields appear in theories of gravitation with extra dimensions, we will consider the radion field which
codes for the size of the compactified space. In four dimensions, it appears as a correction to Einstein's theory and has several
properties in common with the scalar field first introduced for gravitation by Nordstr\"om.

\section{Scalar Gravitation}

The first gravitational theories were inspired by electrostatics. Since Newton's law for the gravitational force 
\beq
         F_N = - G{m_a m_b\over r^2}                                                                 \label{F_N}
\eeq
between two masses $m_a$ and $m_b$ with separation $r$ has exactly the same form as Coulomb's law
\beq
         F_C = {e_a e_b\over 4\pi r^2}
\eeq
for the force between two charges $e_a$ and $e_b$ at the same separation, it has been tempting for more than a hundred years to describe them
in a similar language\cite{Pais}. The only, and crucial difference, is that the gravitational force is always attractive while
the electrostatic force can be both attractive and repulsive, depending on the relative sign of the charges. The r\^ole of
the electromagnetic potential will then be played by a scalar gravitational field. 

\subsection{Non-relativistic interactions}

Let us see how this analogy can be formulated more closely in a Lagrangian description. In electrostatics the electric
field is given in terms of the potential as ${\bf E} = - {\bg\nabla}\phi$. The static Lagrangian
is therefore
\beq
       {\cal L} = {1\over 2}{\bf E}^2 - \rho\phi = {1\over 2} ({\bg\nabla}\phi)^2 - \rho\phi                   \label{L_E}
\eeq
where $\rho(\br) = \sum_a e_a\delta(\br - \br_a)$ is the charge density.
The corresponding equation of motion is just Poisson's equation $\nabla^2\phi = - \rho$. We now find the interaction energy $E$ from
the Hamiltonian $H = \int\!d^3x(- {\cal L})$ which after a partial integration gives
\be
        E = \int\!d^3x\Big(  {1\over 2}\phi\nabla^2\phi +  \rho\phi\Big)
\ee
Using now the Poisson equation in the first term and writing it in momentum space as $\bk^2\phi_\bk = \rho_\bk$ where the 
Fourier-transformed charge density is $\rho_\bk = \sum_a e_a e^{i\bk\cdot\br_a}$, we get back to the Coulomb energy
\beq
        E = {1\over 2}\int\!d^3x \rho\phi\ =  {1\over 2}\sum_\bk {\rho_\bk \rho_{-\bk}\over \bk^2}             \label{interE}
          = \sum_{a < b}{e_a e_b\over 4\pi |\br_a - \br_b|}
\eeq
after throwing away the infinite self-energies.

For the corresponding gravitational interaction between several masses, we can introduce a massless, scalar field $\phi$ described
by the canonical Lagrangian density ${\cal L}_\phi = (1/2)(\del_\mu\phi)^2$. In the static limit we then have for the gravitational field
\beq
         {\cal L} =  - {1\over 8\pi G} ({\bg\nabla}\phi)^2 - \rho\phi                                         \label{L_g1}
\eeq
where now  $\rho(\br) = \sum_a m_a\delta(\br - \br_a)$ is the mass density and $G$ is Newton's gravitational constant entering his
force law (\ref{F_N}). This should be compared with (\ref{L_E}) for the electrostatic case. Notice the sign difference in the kinetic part.
It results in the corresponding field equation 
\beq
           \nabla^2\phi = 4\pi G\rho                                  \label{Poisson}
\eeq
The gravitational interaction energy now follows as in the electrostatic case and becomes
\beq
        E = {1\over 2}\int\!d^3x \rho\phi\ =  -G\sum_{a < b}{m_a m_b\over |\br_a - \br_b|}
\eeq
It is attractive as it should be.

\subsection{Nordstr\"om theory I}

In order to move out of this non-relativistic or static description, such a scalar theory of gravitation must be made consistent
with the special theory of relativity. This was the first problem Nordstr\"om considered\cite{GN_1}. The simplest
generalization of the field equation (\ref{Poisson}) is just
\beq
           \Box\phi = - 4\pi G\rho                                    \label{Poisson_1}                          
\eeq
when written in terms of the d'Alembertian operator $\Box = \del_t^2 - \nabla^2$. This is just what Nordstr\"om first tried. He
could then calculate the gravitational field from masses in arbitrary motion. 

What was needed next, was to find the equation of motion for a particle moving in this external field. This problem required a 
more dramatic answer. Nordstr\"om found that the
inertial mass of the particle was no longer constant and would depend exponentially on the field\cite{Pais}\cite{Norton}. At
that time this was not consider to be problematic and was thought to be similar to its dependence on velocity in the special
theory of relativity. The four-velocity of the particle $u_\mu$ then satisfies the equation of motion
\beq
            \dot{u}_\mu + \dot{\phi}{u_\mu} = \del_\mu\phi                   \label{eom_1}
\eeq 
where the dot derivative is with respect to proper time. All particles will fall the same way in a gravitational field independently of
their masses so the weak equivalence principle is satisfied.

One immediate problem with this theory was that it could not be derived from an action principle. But more importantly, it
was stressed by Einstein that the mass density $\rho$ in (\ref{Poisson_1}) should be a world scalar and preferably the
trace $T$ of the energy-momentum tensor $T_{\mu\nu}$ of the system. The whole question about the relationship between the
inertial and gravitational mass of a particle had to be reconsidered\cite{GN_2}. 

\subsection{Nordstr\"om theory II}

This effort resulted shortly in his second scalar gravitational theory\cite{GN_3}. The mass density $\rho$ was
now assumed to be proportional to the trace $T_m$ of the energy-momentum tensor setting up the gravitational field, i.e. 
$\rho = g(\phi)T_m$ where $g(\phi)$ is some function to be determined. From the requirement of proportionality between inertial and 
gravitational masses, he found  $g(\phi) = 1/\phi$ after a constant shift of the potential\cite{Laue}. 
The field equation (\ref{Poisson_1}) is therefore replaced by 
\beq
           \phi \Box\phi = - 4\pi G T_m                                  \label{Poisson_2}                          
\eeq
and was now non-linear. The electromagnetic field has $T_m = 0$ and it would therefore not have any gravitational interactions. As a
result, there would not be any bending of light near stellar masses in this theory.

The dynamics of a gravitational system of particles could now be derived from an action principle. In particlular, the motion
of a particle in an external field follows from the variational principle
\beq
           \delta\!\int\! ds \phi(x) = 0                                      \label{action}
\eeq
with the line element  $ds = \sqrt{u^\mu u_\mu}d\tau$ where $u_\mu$ is the 4-velocity and $\tau$ the proper time.
Instead of (\ref{eom_1}), it  gives the modified equation of motion
\beq
            \phi\dot{u}_\mu + \dot{\phi}{u_\mu} = \del_\mu\phi                   \label{eom_2}
\eeq 
which obviously also satisfies the weak equivalence principle.

This theory was immediately seized upon by Einstein who in the same year 1913, reformulated it in an elegant way and presented it
as the first consistent, relativistic theory of gravitation\cite{Einstein_2}. It is described by a massless, scalar field $\phi(x)$ with
the standard energy-momentum tensor
\beq
        T_g^{\mu\nu} = {1\over 4\pi G}\Big[\del^\mu\phi\del^\nu\phi - {1\over 2}\eta^{\mu\nu}(\del_\lambda\phi)^2\Big]
\eeq
where $\eta_{\mu\nu} = diag(+1,-1,-1,-1)$ is the Minkowski metric. In addition, matter of density $\rho$ and 4-velocity $u^\mu$ 
contributes 
\beq
        T_m^{\mu\nu} = \rho\phi u^\mu u^\nu                                                    \label{T_mat}
\eeq
so that the trace
$T_m = \rho\phi$. The total energy-momentum tensor $T_g^{\mu\nu} + T_m^{\mu\nu}$ is now conserved. This follows from the divergence
$\del_\mu T_g^{\mu\nu} = \Box\phi(\del^\nu\phi)/4\pi G = -\rho(\del^\nu\phi)$ when we use the field equation of motion (\ref{Poisson_1}).
On the other hand, $\del_\mu T_m^{\mu\nu} = \rho u^\mu\del_\mu(\phi u^\nu)$ when we impose matter conservation $\del_\mu(\rho u^\mu) = 0$.
Since $u^\mu\del_\mu = d/d\tau$, this becomes  $\del_\mu T_m^{\mu\nu} = \rho(\del^\nu\phi)$ from the particle equation of motion 
(\ref{eom_2}). The total divergence therefore adds up to zero and we have conservation of energy and momentum for the whole, interacting system. 

\subsection{Einstein-Fokker geometric reformulation}
At the same time as Nordstr\"om developed this scalar field theory, Einstein was struggling with the development of his own, geometric
formulation of a relativistic theory of gravitation. It was therefore clear to him that the variational principle (\ref{action}) for
the particle motion gives a geodesic equation in a curved spacetime with metric $g_{\mu\nu}(x) = \phi^2(x)\eta_{\mu\nu}$. This represents
a conformal transformation by a factor $\phi^2$ and the resulting spacetime will have a non-zero Ricci curvature tensor. In particular, 
the Ricci scalar is 
\beq
           R = -{6\over\phi^3}\Box\phi
\eeq
as found by Einstein and Fokker\cite{EF}. Under the conformal transformation the trace of the energy-momentum tensor 
will be transformed into $T \ra T/\phi^4$. This can now be used to rewrite the Nordstr\"om gravitational equation (\ref{Poisson_2}) as
\beq
           R = 24\pi G T
\eeq
where on the right-hand side now enters the contributions from both matter and other possible sources. Here we have for the first time
a purely geometric description of gravitational interactions. One can only speculate how much
this equation was on Einstein's mind when he the following year formulated his general theory of relativity where the gravitational tensor 
field equation has exactly the same structure.

\subsection{Scalar theories of gravitation}
One can obviously wonder if the Nordstr\"om gravitational theory is unique. In order to investigate that, let us consider a non-relativistic
particle moving in a gravitational potential $\phi$. It has the Lagrangian $L_p = (1/2)mv^2 - m\phi$. Neglecting relativistic terms of the
order of $\phi v^2$, this follows from the Lorentz-invariant action
\beq
        S_p = -m\int\! ds\,(1 + \phi)
\eeq
where $ds$ is the relativistic line element introduced in (\ref{action}). This can be generalized to
\beq
        S_p = -m\int\! ds\,A(\phi)
\eeq
where the function $A(\phi) \ra \phi + const$ in the weak-field limit $\phi \ra 0$. It results in the equation of motion
\beq
        {d\over d\tau}\big(Au_\mu\big) = A'(\del_\mu\phi)
\eeq
where $A'$ is the derivative of $A$ with respect to $\phi$. If $A = e^\phi$, we recover Nordstr\"om's first particle equation of 
motion (\ref{eom_1}) while $A = \phi$ gives his second equation (\ref{eom_2}).

The Lagrangian for the scalar, gravitational field coupled to particles with a scalar mass density $\rho$, will now be the covariant
generalization of the non-relativistic Lagrangian (\ref{L_g1}) we started out with,
\beq
          {\cal L} =   {1\over 8\pi G} \eta^{\mu\nu}\del_\mu\phi\del_\nu\phi - \rho A(\phi)                     \label{L_g2}
\eeq
It gives the modified field equation $\Box\phi = - 4\pi G\rho A'$. There is no change in the energy-momentum tensor for the field, 
while for the matter part we now have instead $T_m^{\mu\nu} = \rho A u^\mu u^\nu$. The total energy-momentum for the system is conserved
as before. Eliminating the mass density from the field equation using the trace $T_m = \rho A$, we find a generalized Nordstr\"om field equation
\beq
          A(\phi)\Box\phi = -  4\pi GA'(\phi)T_m
\eeq
In the weak-field limit it is equivalent to the previous equations.

As already stated, the choice $A = \phi$ reproduces the second Nordstr\"om theory. Taking instead $A = 1 + \phi$, one obtains a theory
investigated by Bergmann\cite{Bergmann}. The choice $A = e^\phi$ is inspired by his first theory and is given as a home-work problem
in the textbook by Misner, Thorne and Wheeler\cite{MTW}. It has more recently been used by Shapiro and Teukolsky as a laboratory
for numerical, relativistic gravitation\cite{ST}. None of these theories give neither any deflection of light nor the right perihelion
shift for Mercury. 

\section{Extra Dimensions}

There are no experimental indications that our spacetime has any extra dimensions. Experiments in high energy physics
demonstrate that the strong and electroweak particle forces are confined to only four dimensions down to distances of the order of 
$10^{-18}$ m. On the other hand, our understanding of gravitational interactions has only been verified down to separations of the
order of $10^{-4}$ m\cite{Adelberger}. Thus it is conceivable that gravity can extend out into extra dimensions of size smaller than this, 
but still large compared with the range of the electroweak force\cite{ADD}

The possibility that spacetime has more than four dimensions, was first contemplated by Nordstr\"om\cite{GN_4}. This was in order to
unify his scalar theory of gravitation with Maxwell's theory of electromagnetism. Now it turned out that this particular gravitational
theory got a very short lifetime. But just the idea that spacetime has extra dimensions lies today at the heart of the most fundamental
theories of quantum gravity and superstrings. Even if these new dimensions are so microscopically small that they never can be directly
detected, they can still have very profound, indirect consequences in that a complicated and asymmetrical description of the basic
interactions in our four-dimensional spacetime are just the reflections of a highly compact and symmetrical description seen from a
higher-dimensional spacetime.

\subsection{Five-dimensional Maxwell theory}

In addition to constructing a relativistic theory of gravitation, Nordstr\"om wanted also to unify it with electromagnetic 
theory\cite{GN_4}. For this he envisaged our four-dimensional spacetime to be endowed with an extra, fifth dimension. Both of
these interactions should then be part of a Maxwell theory in five dimensions,
\beq
        {\cal L} = -{1\over 4} F_{ab}F^{ab} - J_aA^a
\eeq
where Latin indices now take on five values. The generalized field strengths are  $F_{ab} = \del_aA_b - \del_bA_a$ where the electromagnetic 
vector-potential has the components $A^a = (A^\mu, \phi/\sqrt{4\pi G})$, i.e. incorporating the gravitational potential $\phi$. The corresponding
5-vector for the current densities is $J^a = (J^\mu, \rho\sqrt{4\pi G})$ where $\rho$ is the mass density. From here follows the
5-dimensional wave equation
\beq
         \del^2A_a - \del_a(\del^bA_b) = - J_a                                         \label{5-wave}
\eeq
We now see that ordinary current conservation $\del_\mu J^\mu = 0$ in our spacetime implies that $\rho$ is independent of the coordinate
in the fifth direction, $\del_5\rho = 0$. Nordstr\"om then observed that when the vector-potential is also made independent of this new
coordinate, the 5-dimensional wave equation separates (\ref{5-wave}) into the standard Maxwell wave equation  
$\del^2A_\mu - \del_\mu(\del^\nu A_\nu) = - J_\mu$ plus his wave equation (\ref{Poisson_1}) for the gravitational field. 

This daring and beautiful proposal  was generally overlooked and generated little interest. One reason can be that it was
published at the same time as WW I broke out and not much later Einstein had the correct theory of gravitation\cite{Einstein_1}.

The requirement $\del_5A^a = 0$, that the field should have no variation in the new dimension, was introduced again several years 
later by Kaluza\cite{Kaluza} in his 5-dimensional unified theory of gravitation and electromagnetism built upon Einstein's theory. It 
is now called the cylinder condition and was explained by Klein\cite{Klein} as a consequence of quantum mechanics when the extension in 
the extra dimension is so small that it is unobservable.

\subsection{Einstein gravitation with extra dimensions}

In order to illustrate some of the physics resulting from having more than four dimensions, let us with Kaluza and Klein consider Einstein 
gravity in a spacetime with $D = 4 + n$. If the extra dimensions have the coordinates $y^\alpha$, the fundamental Einstein-Hilbert action will
be
\beq
        S = -{1\over 2}M_D^{2+n}\int\!d^4x\int\!d^ny\sqrt{-\bar{g}}\,\bar{R}                                    \label{EH_D}
\eeq
where $M_D$ is the Planck mass in this spacetime. It has the metric $\bar{g}_{ab}$ with determinant $\bar{g}$ and $\bar{R}$ is the
corresponding Ricci curvature scalar. Assuming that the extra dimensions are compactified and described by the internal metric 
$h_{\alpha\beta}$, we assume the form 
\beq
         d\bar{s}^2 = g_{\mu\nu}(x)dx^\mu dx^\nu - b^2(x)h_{\alpha\beta}(y)dy^\alpha dy^\beta
\eeq
for the full spacetime metric in its groundstate. The unknown function $b(x)$ gives the general size of the compactified volume 
$V_n = \int\!d^ny\sqrt{-\bar{h}}$. With the full metric on this form, one can calculate the scalar curvature $\bar{R}$.  Since the determinant 
$\bar{g} = gh$ now separates into a product, the full action can be written as
\beq
       S = - {1\over 2}M_D^{2+n}V_n\int\!d^4x\sqrt{-g}b^n\Big[R - {1\over b^2}R_n + n(n-1)g^{\mu\nu}\del_\mu b\del_\nu b\Big]\label{extra}
\eeq
after a few partial integrations and assuming the the internal curvature scalar $R_n$ is a constant\cite{Petter}. The coefficient in front
of the four-dimensional Ricci scalar $R$ defines the usual Planck constant
\beq
        M_P^2 = V_nM_D^{2+n}
\eeq
in agreement with more general considerations\cite{ADD}. With two extra dimensions and the requirement that the gravitational range in these
directions must not be larger than  $10^{-4}$ m, one finds a value for the mass $M_{D=6}$ which can be as small as 1 TeV. In this case one
would have the possibility to see the effects of quantum gravity or string theory at LHC and other future high-energy particle accelerators.

The result (\ref{extra}) for the action describes a scalar field $b(x)$ with gravitational interactions. It can be rewritten on a more
canonical form by defining a new field $\Phi = b^n$ which then gives
\beq
     S =  - {1\over 2} M_P^2\int\!d^4x\sqrt{-g}\Big[\Phi R + (1-1/n){1\over\Phi}g^{\mu\nu}\del_\mu\Phi\del_\nu\Phi 
          - R_n\Phi^{1-2/n}\Big]                      \label{BD}
\eeq
This extended tensor theory is now recognized to be of the Brans-Dicke type\cite{BD} where the last term is a potential $V(\Phi)$ 
for the scalar field 
resulting from the curvature of the internal space. Without such a potential, it gives corrections to standard Einstein gravity which
depends on the parameter $\omega = -(1-1/n)$ in front of the scalar kinetic energy. When $n>1$ we will have $\omega$ to be close to one in 
magnitude and negative. On the other hand, solar system gravitational precision measurements require $\omega > 3500$\cite{Will}. This
doesn't mean that such scalar-tensor theories are ruled out by observations. Damour and Nordtvedt have shown that if $\omega$ gets
to be dependent on $\Phi$ which easily happens when higher order corrections are included, it will be driven towards very high values today
because of the cosmological evolution of the Universe\cite{DN}. 

\subsection{Radions and quintessence}

In this extra-dimensional approach the Brans-Dicke scalar field $\Phi(x)$ stems from the radial size $b(x)$ of the compactified space.
For this reason it is called a radion and is generic in this kind of theories. It plays an important r\^ole in todays cosmological
theories\cite{Kolb}, and is used both for inflation\cite{Mazumdar} and quintessence in the Universe\cite{WR}. These physical 
phenomena are usually discussed in the Einstein frame which arises from a conformal transformation $g_{\mu\nu} \ra A^2(\Phi)g_{\mu\nu}$ 
of the metric in
the Brans-Dicke action (\ref{BD}) such that the cofficient of the Einstein-Hilbert term $R$ is $\Phi$-independent and with the  
value one. This comes about since this term will transform as $R \ra A^{-2}R - 6A^{-3}\Box A$ where now the $\Box$ operator is the
d'Alembertian in curved space with metric $g_{\mu\nu}$. This is the same conformal transformation as used by Einstein and Fokker in
their reformulation of the Nordstr\"om scalar theory, but then from flat Minkowski space. One obtains the desired result with the
choice $A = \Phi^{-1/2}$ which now will appear as the new scalar field in the Lagrangian. 

In order to get the kinetic term on canonical form, we let $A = A(\phi)$ satisfy
\beq
           {1\over A}{\del A\over\del\phi} = - {\alpha\over M_P}
\eeq
where the parameter $\alpha$ is related to the Brans-Dicke parameter $\omega$ by
\beq
             \alpha^2 = {1\over 4\omega + 6}
\eeq
Since this is constant, we thus simply have $A = e^{-\alpha\phi/M_p}$. The resulting action in the  Einstein frame is therefore
\beq
            S =  \int\!d^4x\sqrt{-g}\Big[-{1\over 2} M_P^2 R + {1\over 2}g^{\mu\nu}\del_\mu\phi\del_\nu\phi - \tilde{V}(\phi)\Big]
\eeq  
where $\tilde{V}$ is the transformed potential, 
\beq
        \tilde{V}(\phi) =  {1\over 2} M_P^2A^4(\phi)V(\Phi)
\eeq   
If the original potential $V$ is just a constant, as would be the case if the compact space was flat and we instead had included a
cosmological constant in the $D$-dimensional action (\ref{EH_D}), the resulting potential in the Einstein frame is seen to be an
exponential. Such a potential has many attractive features which today is used in realistic quintessence models\cite{quint}.

Specific models depend on the number $n$ of extra dimensions and what kind of fields are included in the fundamental action. A 
more detailed investigation of such a model with $n=2$ and an additional scalar field in the extra-dimensional spacetime has
been performed by Albrecht, Burgess, Ravndal and Skordis\cite{ABRS}. Due to quantum effects, the Brans-Dicke parameter will be field
dependent and thus evolve into a sufficiently large value today so to be consisent with solar system gravity tests. In addition, the
same quantum effects give a radiative correction to the purely exponential potential which then takes the more realistic form
previously proposed by Albrecht and Skordis\cite{AS}. 

The radion field also couples to matter. In the original frame this would have been described by an action $S_m = S_m[g_{\mu\nu},\psi]$
where $\psi$ stands for a generic set of matter fields. Under the conformal transformation it changes into 
$S_m \ra S_m[ A^2(\phi) g_{\mu\nu},\psi]$ which means that the radion field will modify the fundamental coupling constants in the matter
Lagrangian. Since $\phi$ varies with the evolution of the Universe, this means that the fundamental constants also will vary with time.
There are some indications that this might actually be the case for the electromagnetic fine-structure constant\cite{alpha}.

From the above it is seen that the coupling of the radion to matter is only through the factor $A^2$ in the metric. This coupling will
then manifest itself in the equation of motion for the radion which now becomes\cite{Petter}
\beq
        \Box\phi + \tilde{V}'(\phi) = {\alpha\over M_P}T_m
\eeq
The Brans-Dicke parameter $\alpha$  gives the deviation of the coupling strength from that of the graviton. We notice that this wave
equation is of the same form as the Nordstr\"om scalar field equation when the potential is zero.

\section{Conclusion}

Scalar fields are abundant in modern theories of gravitation and other fundamental interactions. It was Nordstr\"om who first
constructed such a consistent, relativistic theory based on a single field. Although his scalar theory was shortly afterwards
replaced by Einstein's tensor theory, scalar fields in this connection are today again being investigated and with greater vigor 
than ever before.
There are many experimental indications from physics and cosmology that they are present in our physical world, but it requires much more 
work to find out how they fit in with the other fundamental fields we know.

The intellectual leap he made suggesting that our spacetime has extra dimensions in order to unify the different interactions we see around
us, has today resulted in a large number of very promising theoretical proposals and experimental efforts. It will probably never be possible
to rule out such an extension of our conventional spacetime. One can instead only hope that their presence in some way soon will be
demonstrated and bring order and beauty to our understanding of the physical world.

{\bf Acknowledgement:} I want to thank Kari Enqvist and other members of the HIP cosmology group at the University of Helsinki for 
providing an inspirational environment in which this contribution was conceived. In addition, I'm grateful to Christofer 
Cronstr\"om and other members of the organizing committee for inviting  me to the first Gunnar Nordstr\"om Symposium. 
This work has been supported by NorFA.

\end{document}